# The origins of the leakage currents of p-n junction and Schottky diodes in all kinds of materials: A novel explanation based on impurity-photovoltaic-effect due to the self-absorption of the room-temperature infrared emission from materials

Jianming Li

Institute of Semiconductors, Chinese Academy of Sciences, A35 Qinghua East Road, Haidian District, Beijing 100083, P. R. China

**ABSTRACT**

A p-n junction is the basic building block for various semiconductor devices. A Schottky diode has characteristics that are essentially similar to those of the p-n junction diode. As is known, the leakage currents of p-n and Schottky junctions affect the overall performance of the devices and reduce the reliability of the devices. In order to achieve optimum device design, it is essential to fully understand the physical principle of the leakage currents. In traditional theory, defects provide a path for leakage current to travel. In this study, a novel theoretical model based on impurity-photovoltaic-effect is proposed to explain the leakage currents. It is well known that any object at a room-temperature emits infrared (IR) photons due to blackbody radiation. As is also well known, there is no absolutely pure material, and any material contains unavoidable defects associated with impurities. The self-absorption of the IR emission could be achieved through the sub-band-gap excitations due to defect-related intermediate levels in forbidden energy band-gap, creating carriers (electrons and holes). Some of the carriers diffuse into the built-in electric field of the junction. The built-in field then sweeps out electrons and holes in opposite directions, forming IR-generated photocurrent. Therefore, the leakage current is regarded as photocurrent. In addition to p-n junctions, some junctions exist in many semiconductor devices such as p-i-n diode and charge-coupled device (CCD), and these junctions also have built-in field due to contact potential difference. In fact, every semiconductor device contains at least one junction. The novel model is expected to explain the leakage current for all kinds of junctions with semiconductor built-in fields.

The p-n junctions are of great importance in modern electronic applications. Most semiconductor devices contain at least one p-n junction. The p-n junctions are currently widely used as devices in applications such as rectifiers, lasers, light-emitting diodes (LEDs), bipolar transistors, various field-effect transistors (FETs), solar cells, photodetectors, multipliers, etc. Especially, a very large-scale integration (VLSI) chip contains a huge number of p-n junctions.

A Schottky diode is electrically similar to the abrupt one-sided p-n junction diode, but Schottky devices have a high switching speed and are suitable for high-speed applications such as digital switches and microwave detectors, etc.

The p-n junction theory serves as the foundation of the physics of semiconductor devices and as the guidance for structural design, material selection, and process sequence control of device fabrication.

The diodes were discovered about a hundred years ago. The basic theory of current-voltage characteristics of the p-n junction diode was established by Shockley [1]. The Shockley theory adequately predicts the current-voltage characteristics of germanium p-n junction at low current densities. For Si and GaAs p-n junctions, however, the theory can give only qualitative agreement. [2] Thus, the theory was subsequently developed by Sah *et al.* [3] and Moll [4].

In a real diode, a number of sources may lead to bandgap states, i.e., there are defects and associated bandgap states that may lead to trapping, recombination, or generation terms. The states may arise if the material quality is not very pure so that there are chemical impurities present. The doping process itself can cause defects such as vacancies, interstitials, etc.

Material-defects indeed can cause undesirable reverse leakage currents that are not predicted by the ideal diode characteristics [5]. Such leakage currents have been widely observed in experimental measurements.

The leakage currents are considered an important factor affecting device performance such as reliability, power consumption, efficiency, resolution, noise, etc. As the integration scale of VLSIs increases, degradation of device characteristics due to p-n junction leakage is becoming a serious problem. The refresh operation failure in Dynamic Random Access Memory (DRAM) cells due to large leakage current is one typical example. In order to achieve optimum device performance, it is essential to fully understand the origin and physical mechanisms of the leakage currents.

A number of attempts have been made to model reverse currents of p-n junctions in various real semiconductor materials, employing mechanisms such as Pool-Frenkel (PF) emission [6-8], variable range hopping (VRH) [9-15], trap-assisted tunneling (TAT) [16,17], thermionic-field emission (TFE) [18-20], phonon-assisted tunnelling (PhaT) [21], and thermionic emission (TE) based on Boltzmann statistics [22], etc.

Many studies on the leakage mechanisms indicate that impurity-related defects provide a leakage path for current to travel. In this study, an impurity-photovoltaic-effect based model explaining the leakage currents is proposed as follows.

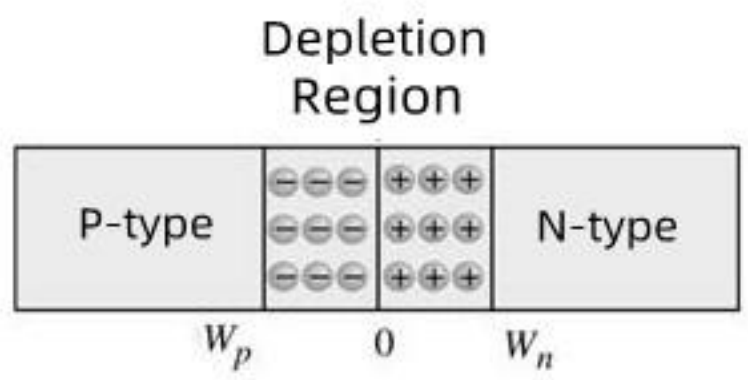

Fig. 1. A schematic of a p-n junction without bias, showing the depletion regions.

First, let us consider a zero-bias p-n junction diode under illumination. The p-n junction between p-type and n-type material has a built-in space-charge layer or depletion region, as shown in Fig. 1. Light that enters the diode and reaches the depletion region of the p-n junction generates electron–hole pairs (EHPs). The built-in electric field of the depletion region sweeps out electrons and holes in opposite directions, creating photocurrent. It is also possible for EHPs to be generated within about one diffusion length on either side of the depletion region and through diffusion to reach the depletion region, making a contribution to the photocurrent. This is well-known photovoltaic effect.

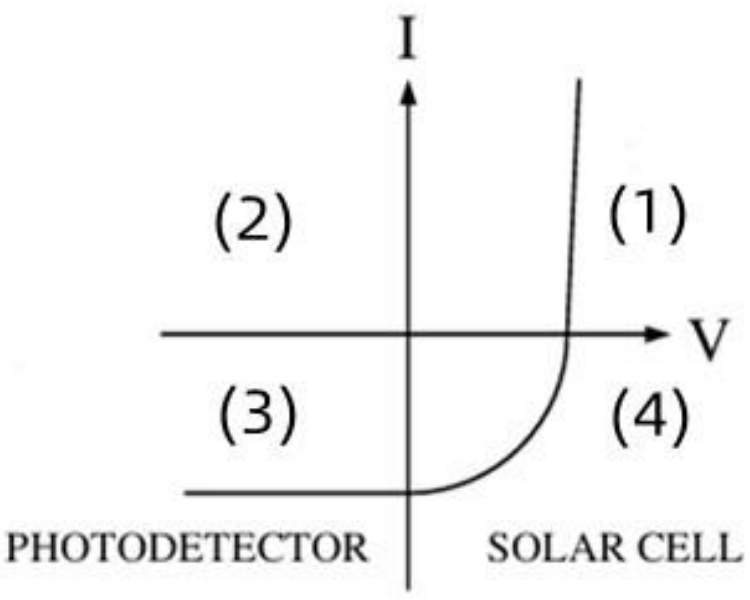

Fig. 2. Current-voltage plot showing regimes of operation for a p-n diode under illumination.

Figure 2 is current-voltage (I-V) plot showing regimes of operation for a p-n junction diode under illumination. When operated in quadrant (3), the device acts as a photodetector, whereas in quadrant (4) it behaves as a solar cell. When the applied voltage equals zero, the current of the device is just the short circuit current of solar cell. The diode with zero bias has the properties of a current source (equivalently a current with voltage).

For a p-n junction diode without illumination, the p-n junction is still subjected to photon irradiation. This can be understood as follows. As is well known, the vibratory motion of the electrical charges in an object containing atoms (composed of electrons and atomic nucleuses) leads to thermal radiation, which is called blackbody radiation, and any object at a room-temperature emits infrared (IR) photons. An experimental result [23] supports that semiconductor material at room temperature emits IR photons. The IR photons that are emitted within the diode could be absorbed before the photons leave the diode, and the self-absorption is achieved under certain conditions.

In a real p-n diode, the defects associated with different impurities create intermediate levels (bandgap states) in forbidden energy band-gap. Furthermore, the self-absorption of the IR emission within the diode could be achieved through the sub-band-gap excitations shown in Fig. 3, which produces carriers. Some of the carriers diffuse into the built-in field of the p-n junction and subsequently drift across the depletion region, creating IR-generated current.

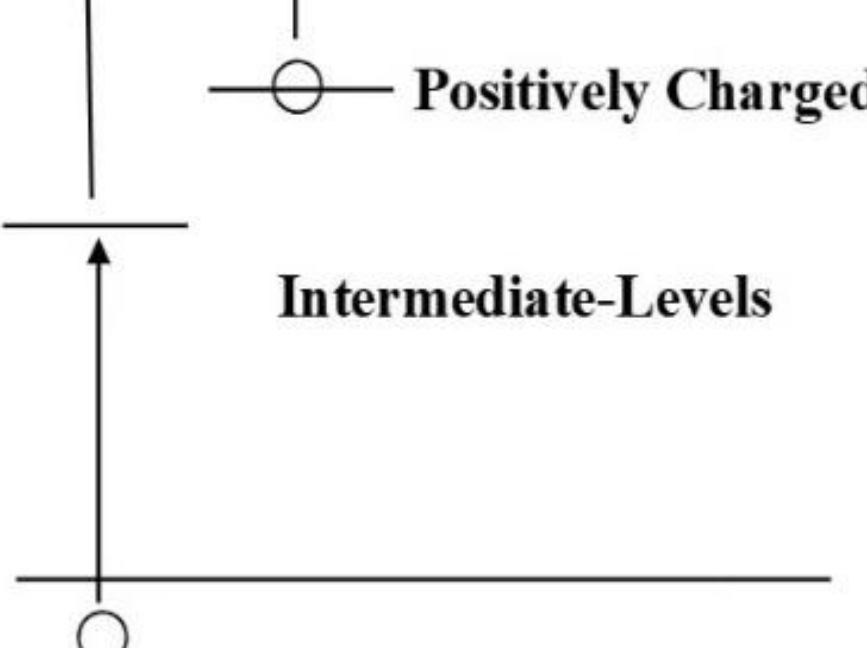

Fig. 3. The sub-band-gap excitation mechanism via the intermediate levels within forbidden gap.

According to the impurity-photovoltaic-effect, the reverse saturation current of a p-n junction diode can be regarded as photocurrent. In other word, the leakage current of a diode is attributed to photocurrent resulting from the self-absorption of the IR emission within the diode.

Many studies show that operating temperature has a profound effect on p-n junctions. In the reverse-bias conditions, the saturation current increases with temperature. [24] This can be explained in terms of impurity-photovoltaic-effect because the intensity of the IR emission within a diode increases with temperature.

For a p-n junction diode under reverse bias, the applied voltage expands the width of the space-charge (depletion) region, but carrier diffusion length on either side of the depletion region remains unchanged. As a result, the reverse voltage leads to carrier creation in wider range. If the diode contains certain defects, the reverse leakage current obviously increases with voltage. Therefore, it can be explained that the leakage has strong electric field dependence. [25]

An experimental result shows that the surface carrier generation makes a significant contribution to the reverse leakage current of a p-n junction. [26] This phenomenon can be explained as follows. The surface of a p-n junction diode has numerous defects (lattice mismatch dislocation). Such defects enable the self-absorption of the IR emission to occur through the sub-band-gap excitations, creating photo-generated carriers. The carriers could diffuse into the depletion region of the p-n junction. Furthermore, the carriers could be collected as photocurrent. Therefore, the surface region could be believed to act as an important source, generating carriers which could diffuse into the built-in field of the p-n junction. It is thus clear that the effect of surface carrier generation on junction leakage can be explained using the impurity-photovoltaic-effect based model.

As is reported [27], reverse leakage currents were observed in GaN p–n junctions nearly free of dislocations. This means that a few of defects still cause photocurrent to occur according to impurity-photovoltaic-effect.

Another study [28] implies that a p-n junction in very pure silicon creates IR-generated current through impurity-photovoltaic-effect, even though the silicon device contains few impurities. Among all kinds of semiconductor materials, the silicon material used in the work described in Ref. 28 is believed to contain fewer defects than any other material.

Accordingly, the model based on impurity-photovoltaic-effect is proposed to interpret the leakage currents of the p-n junctions in all kinds of semiconductor material.

In fact, defects associated with different impurities are always present in any semiconductor material because there is no absolutely pure material. Each kind of semiconductor materials has its own characteristic defects such as intrinsic defects with self-owned intermediate levels in forbidden energy band-gap, etc. Therefore, the reverse leakage currents of p-n junctions in various materials have different characteristics. According to this study, various reverse leakage current phenomena can be explained in terms of impurity-photovoltaic-effect.

In addition to p-n junctions, some junctions exist in many semiconductor devices such as p-i-n diode and charge-coupled device (CCD), and these junctions also have built-in field due to contact potential difference. In fact, every semiconductor device contains at least one junction. The impurity-photovoltaic-effect based model is expected to explain the leakage current for all kinds of junctions with semiconductor built-in fields.

## DECLARATION OF COMPETING INTEREST

This is to certify that this manuscript does not make any conflict of interest with any person or institution or laboratories or any work.

## ACKNOWLEDGEMENT

This work was financially supported by Chinese Academy of Sciences (CAS).

## REFERENCES

[1] W. Shockley, "The theory of p-n junction in semiconductors and p-n junction transistors," *Bell Syst. Tech. J.*, vol. 28, p. 435, 1949.

[2] S. M. Sze, *Physics of Semiconductor Devices*. New York, USA: John Willey & Sons, 1981, pp. 89-90.

[3] C. T. Sah, R. N. Noyce, and W. Shockley, "Carrier generation and recombination in p-n junction and p-n junction characteristics," *Proc. IRE*, vol. 45, p. 1228, 1957.

[4] J. L. Moll, "The evolution of the theory of the current-voltage characteristics of p-n junctions," *Proc. IRE*, vol. 46, p. 1076, 1958.

[5] U. K. Mishra and J. Singh, *Semiconductor Device Physics and Design*. Dordrecht, Netherlands: Springer, 2008, p. 170.

[6] J. Frenkel, "On Pre-Breakdown Phenomena in Insulators and Electronic Semi-Conductors," *Phys. Rev.*, vol. 54, pp. 647-648, 1938. https://doi.org/10.1103/PhysRev.54.647

[7] R. M. Hill, "Poole-Frenkel conduction in amorphous solids," *Philos. Mag.*, vol. 23, pp. 59-86, 1971. https://doi.org/10.1080/14786437108216365

[8] K. Fu, H. Fu, X. Huang, T. Yang, C. Cheng, P. R. Peri, H. Chen, J. Montes, C. Yang, J. Zhou, X. Deng, X. Qi, D. J. Smith, S. M. Goodnick, and Y. Zhao, "Reverse Leakage Analysis for As-Grown and Regrown Vertical GaN-on-GaN Schottky Barrier Diodes," *IEEE J. Electron Devices Soc.* vol. 8, pp. 74-83, 2020. https://doi.org/10.1109/JEDS.2020.2963902

[9] R. M. Hill, “Hopping conduction in amorphous solids,” *Philos. Mag.*, vol. 24, pp. 1307-1325, 1971. https://doi.org/10.1080/14786437108217414

[10]H. Iwano, S. Zaima, and Y. Yasuda, “Hopping conduction and localized states in p-Si wires formed by focused ion beam implantations,” *J. Vacuum Sci. Technol. B Microelectron. Nanometer Structures Process. Meas. Phenom.*, vol. 16, pp. 2551–2554, 1998. https://doi.org/10.1116/1.590208

[11]D. V. Kuksenkov, H. Temkin, A. Osinsky, R. Gaska, and M. A. Khan, “Origin of conductivity and low-frequency noise in reverse-biased GaN p-n junction,” *Appl. Phys. Lett.*, vol. 72, no. 11, pp. 1365-1367, 1998. https://doi.org/10.1063/1.121056

[12]D. Yu, C. Wang, B. L. Wehrenberg, and P. Guyot-Sionnest, “Variable range hopping conduction in semiconductor nanocrystal solids,” *Phys. Rev. Lett.*, vol. 92, no.21, 2004, Art. no. 216802. https://doi.org/10.1103/PhysRevLett.92.216802

[13]P. Achatz, O. A. Williams, P. Bruno, D. M. Gruen, J. A. Garrido, and M. Stutzmann, “Effect of nitrogen on the electronic properties of ultrananocrystalline diamond thin films grown on quartz and diamond substrates,” *Phys. Rev. B*, vol. 74, 2006, Art. no. 155429. https://doi.org/10.1103/PhysRevB.74.155429

[14]Q. Shan, D. S. Meyaard, Q. Dai, J. Cho, E. F. Schubert, J. K. Son, and C. Sone, “Transport-mechanism analysis of the reverse leakage current in GaInN light-emitting diodes,” *Appl. Phys. Lett.*, Vol. 99, 2011, Art. no. 253506. https://doi.org/10.1063/1.3668104

[15]G. Greco, P. Fiorenza, M. Spera, F. Giannazzo, and F. Roccaforte, “Forward and reverse current transport mechanisms in tungsten carbide Schottky contacts on AlGaN/GaN heterostructures,” *J. Appl. Phys.*, vol. 129, 2021, Art. no. 234501. https://doi.org/10.1063/5.0052079

[16]M. P. Houng, Y. H. Wang, and W. J. Chang, “Current transport mechanism in trapped oxides: A generalized trap-assisted tunneling model,” *J. Appl. Phys.*, vol. 86, no. 3, pp. 1488–1491, 1999. https://doi.org/10.1063/1.370918

[17]Z. H. Liu, G. I. Ng, S. Arulkumaran, Y. K. T. Maung, and H. Zhou, “Temperature-dependent forward gate current transport in atomic-layer-deposited $Al_2O_3$/AlGaN/GaN metal-insulator-semiconductor high electron mobility transistor,” *Appl. Phys. Lett.*, vol. 98, no. 16, 2011, Art. no. 163501. https://doi.org/10.1063/1.3573794

[18]M. Higashiwaki, K. Konishi, K. Sasaki, K. Goto, K. Nomura, Q. T. Thieu, R. Togashi, H. Murakami, Y. Kumagai, B. Monemar, A. Koukitu, A. Kuramata, and S. Yamakoshi, “Temperature-dependent capacitance–voltage and current–voltage characteristics of Pt/Ga2O3 (001) Schottky barrier diodes fabricated on n—$Ga_2O_3$ drift layers grown by halide vapor phase epitaxy,” *Appl. Phys. Lett.*, vol. 108, 2016, Art. no. 133503. http://dx.doi.org/10.1063/1.4945267

[19]B. Wang, M. Xiao, X. Yan, H. Y. Wong, J. Ma, K. Sasaki, H. Wang, and Y. Zhang, “High-voltage vertical $Ga_2O_3$ power rectifiers operational at high temperatures up to 600K,” *Appl. Phys. Lett.*, vol. 115, 2019, Art. no. 263503. ttps://doi.org/10.1063/1.5132818

[20]F. A. Padovani and R. Stratton, “Field and thermionic-field emission in Schottky barriers,” *Solid State Electron.* Vol. 9, no. 7, pp. 695-707, 1966. https://doi.org/10.1016/0038-1101(66)90097-9

[21]G. Vincent, A. Chantre, and D. Bois, “Electric field effect on the thermal emission of traps in semiconductor junctions,” *J. Appl. Phys.* vol. 50, no. 8, pp. 5484–5487, 1979. https://doi.org/10.1063/1.326601

[22]S. M. Sze and K. K. Ng, *Physics of Semiconductor Devices*. 3rd ed. Hoboken, NJ, USA: John Willey & Sons, 2007, Chapter 2.

[23]J. Li, “A photovoltaic effect with external quantum efficiency above 100%: A support to the explanation for the non-traditional technique raising efficiencies for all kinds of solar cells,” *2025 IEEE 53rd Photovoltaic Specialists Conference (PVSC)*. https://ieeexplore.ieee.org/document/11132573 *arXiv*:2409.20066. https://arxiv.org/abs/2409.20066

[24]E. J. Miller, E. T. Yu, P. Waltereit, and J. S. Speck, “Analysis of reverse-bias leakage current mechanisms in GaN grown by molecular-beam epitaxy,” *Appl. Phys. Lett.*, vol. 84, no. 4, pp. 535-537, 2004. http://dx.doi.org/10.1063/1.1644029

[25]K. Ohyu, M. Ohkura, A. Hiraiwa, and K. Watanabe, “A mechanism and a reduction technique for large reverse leakage current in p-n junctions,” *IEEE Transactions on Electron Devices*, Vol. 42, No. 8, pp. 1404-1412, 1995. https://doi.org/10.1109/16.398655

[26]C. J. Kircher, “Comparison of leakage currents in ion-implanted and diffused p-n junctions,” *J. Appl. Phys.* Vol. 46, no. 5, pp. 2167–2173, 1975. https://doi.org/10.1063/1.321860

[27]T. Shoji, T. Narita, Y. Nagasato, M. Kanechika, T. Kondo, T. Uesugi, K. Tomita, S. Ikeda, T. Mori, and S. Yamaguchi, “Analysis of intrinsic reverse leakage current resulting from band-to-band tunneling in dislocation-free GaN p–n junctions,” *Appl. Phys. Express*, vol. 14, 2021, Art. no. 114001. https://doi.org/10.35848/1882-0786/ac2a03

[28]J. Li, “A non-traditional technique raising efficiencies for all kinds of solar cells,” *J. Environ. Sci. Eng. A*, vol. 12, pp. 214-222, 2023. https://doi.org/10.17265/2162-5298/2023.06.002

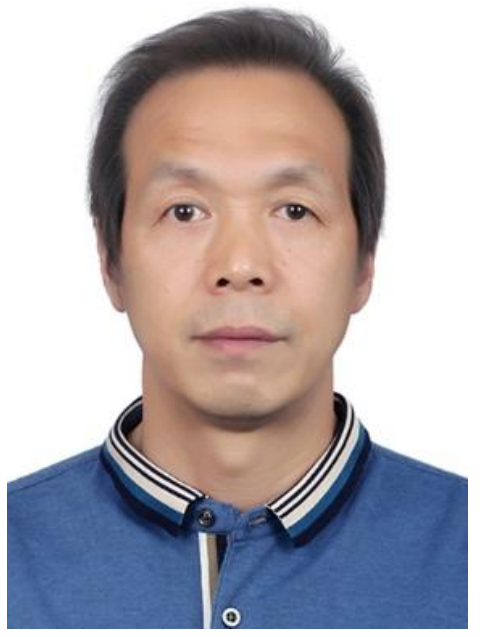

**Jianming Li**, research associate and senior engineer at the Institute of Semiconductors, Chinese Academy of Sciences. Email: jml_iscas@163.com; Tel.: +86-18310559676

Jianming Li was born in Beijing, China, in 1961. He studied at Beijing Normal University from 1980 to 1987. After graduated from the University, he joined the Institute of Semiconductors, Chinese Academy of Sciences. Since 1987, he has been engaged in research on semiconductor materials and photovoltaics, etc. As a Visiting Scientist, he worked at Brookhaven National Laboratory, USA, from 1992 to 1994. As a Visiting Scientist, he worked at Pennsylvania State University, USA in 1997.